\documentclass[prl, aps, 12pt, showkeys, showpacs,epsfig,axodraw]{revtex4}
\usepackage{graphicx}
\begin{document}

\title{Sea Quark Flavor Asymmetry of Hadrons in Statistical Balance Model  }

\author{Bin Zhang}
\email{zb@mail.tsinghua.edu.cn(Communication author)}
 \affiliation{ Department of Physics,
Tsinghua University, Beijing, 100084, China} \affiliation{Center for
High Energy Physics, Tsinghua University, Beijing, 100084, China}
\author{Yong-Jun Zhang}
\email{yong.j.zhang@gmail.com} \affiliation{Science College,
Liaoning Technical University, Fuxin, Liaoning 123000, China}

\begin{abstract}
  We suggested a Monte Carlo approach to simulate a kinetic equilibrium
  ensemble, and proved the equivalence to the linear equations method
on equilibrium.  With the convenience of the numerical method, we
introduced variable splitting rates representing the details of the
dynamics as model parameters which were not considered in previous
works. The dependence on model parameters was studied, and it was
found that the sea quark flavor asymmetry weakly depends on model
parameters. It reflects the statistics principle contributes the
dominant part of the asymmetry and the effect caused by details of
the dynamics is small. We also applied the Monte Carlo approach of
the statistical model to predict the theoretical sea quark
asymmetries in kaons, octet baryons $\Sigma$, $\Xi$, and $\Delta$
baryons, even in exotic pentaquark states.

\end{abstract}


 \maketitle

\section{sea-quark flavor asymmetry from statistical balance model}
Although the proton is the simplest system in which the three colors
of QCD  neutralize into a colorless bound state, we still do not
know how to describe the proton in terms of its fundamental quark
and gluon degrees of freedom from basic principles. The structure of
the proton is rather complicated due to the nonperturbative and
relativistic nature of the quark and gluon in the protons. The
complication also comes from the presence of sea quarks in the
proton. The sea flavor symmetry naively assumed in the Gottfried sum
rule~\cite{Gottfried},which is a symmetry between the light flavor
$u$ and $d$ sea quarks inside the proton, was disproved by
experiments of both deep inelastic scattering and Drell-Yan
processes \cite{FlavorAsymmetry,nmc,NA51,HERMES,E866a,E866}.

Many theoretical attempts have been made to describe the origin of
the nucleon sea and its antiquark asymmetry
~\cite{FlavorAsymmetry,Speth,Kumano,Melnitchouk,Nikolaev,Henley,Magnin,Garvey,Pasquini,Eichten,Cheng,Brown,Ding,Pobylitsa,Zhang1}.
It is assumed that the primary mechanism to generate the sea is
gluon splitting into $u{\bar u}$ and $d{\bar d}$ pairs. Field and
Feynman \cite{Fiel77} suggested that the extra valence $u$ quark in
the proton could lead to a suppression of $g \rightarrow u{\bar u}$
relative to $g \rightarrow d{\bar d}$ via Pauli blocking. But a
subsequent calculation \cite{Ross79} found that the effects of Pauli
blocking are very small, and this result has been confirmed by
another calculation\cite{Stef97}. Thus, it is believed that there
must be a nonperturbative origin. For example, the meson-cloud
inside the nucleon can account for such
asymmetry~\cite{FlavorAsymmetry,Speth,Kumano,Melnitchouk,Nikolaev,Henley,Magnin,Garvey,Pasquini}
and chiral quark models~\cite{Eichten,Cheng,Brown,Ding}. Also the
large-$N_c$ approach~\cite{Pobylitsa} can explain the flavor
asymmetry of the antiquark distribution.

Another attempt to understand the sea flavor asymmetry of the proton
is from a pure statistical consideration in a kinetic equilibrium
model~\cite{Zhang1} or ``statistical balance model'' as called in
previous papers. The idea is rather simple and perspicuous: while
the sea quark-antiquark $u\bar{u}$ and $d\bar{d}$ pairs can be
produced by gluon splitting with equal probabilities, the
time-reversal invariant processes of the annihilation of the
antiquarks with their quark partners into gluons are not flavor
symmetric due to the net excess of $u$ quarks over $d$ quarks. As a
consequence, the $\bar{u}$ quarks have a larger probability to
annihilate with the $u$ quarks than that of the $\bar{d}$ quarks,
and this brings an excess of $\bar{d}$ over $\bar{u}$ inside the
proton. Taking the proton as an ensemble of a complete set of
quark-gluon Fock states, and assuming the probability of `arriving
in' one state from others equals to the probability of `leaving' it,
one can obtain the probabilities of finding every Fock state (state
density) in the proton. Thus one can calculate the quark and gluon
content of the nucleon from a pure statistical consideration. It is
interesting that the model gives a sea flavor $\bar{u}$ and
$\bar{d}$ asymmetry as $[\bar{d}-\bar{u}] \sim 0.132$, which agrees
with the experimental data

The following diagram can describe the `state shifting' between
states.
\begin{figure}[h]
\center
\includegraphics[width=6cm]{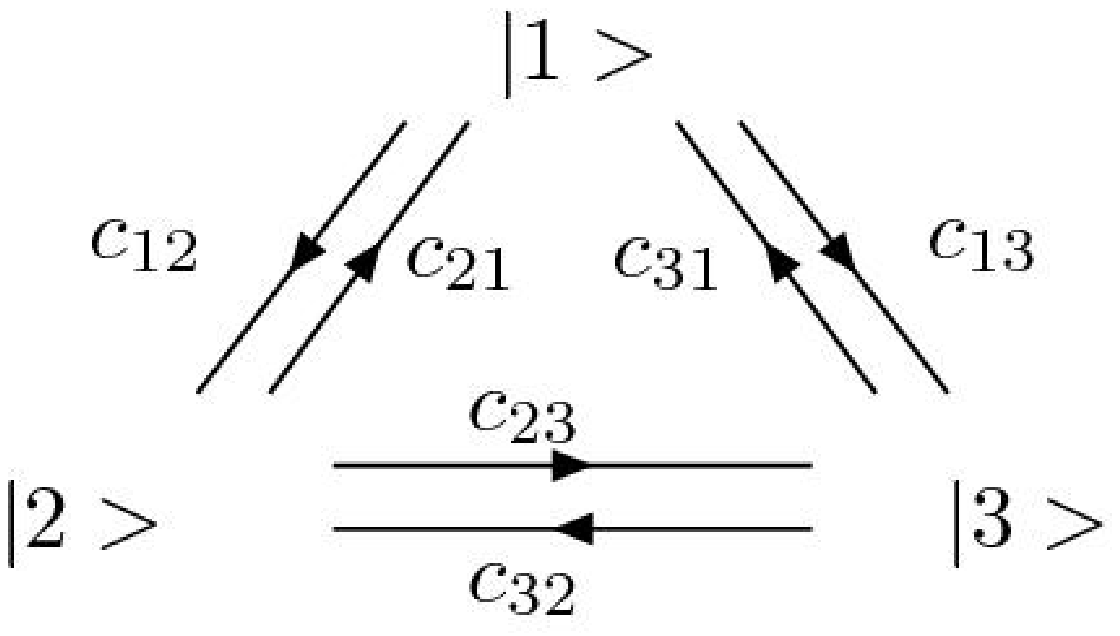}
\end{figure}



Assuming kinetic equilibrium, we have these kinetic equilibrium
equations:
\begin{eqnarray}\label{eq1}
\sum_{j\neq i}^n c_{ij}\rho_{i}=\sum_{j \neq i}^n c_{ji}\rho_{j},
\end{eqnarray}
where $\rho_{i}$ is $|i>$ state density, $c_{ij}$ is the
non-normalized state-shift probability(NSSP) of $ |i>\rightarrow|j>
$,$n$ is the total state number. Also there is the normalization
condition
\begin{eqnarray}\label{eq2}
 \sum_{i}^n\rho_{i}=1.
\end{eqnarray}
If we know $c_{ij}$, we can derive state densities $\rho_{i}$'s by
solving a system of $n$ linear algebraic equations when $n$ is a
finite number. If $n$ is infinite, we can get $\rho_{i}$ by
asymptotic approach in some case if $\rho_{i}$ converges as
$n\rightarrow\infty$. Actually, if we change  $c_{ij}$ to
$c_{ij}/C_0$, where $C_0$ is a arbitrary constant, the result would
be the same. It means we only need the ratios of NSSPs $c_{ij}$'s.

If only considering the particle numbers of quark, anti-quark and
gluon, the proton state can be described as an ensemble of Fock
states
\begin{eqnarray}
&&|uud>,|uudg>,|uudu\bar{u}>,|uudd\bar{d}>,|uudd\bar{d}>,\cdots
\nonumber\\
&&\cdots, |N_u,N_d,N_{\bar{u}},N_{\bar{d}},N_{g}> , \cdots \nonumber
\end{eqnarray}
Because the  u-quark number $N_u\equiv N_{\bar{u}}+2$,and $N_d\equiv
N_{\bar{d}}+1$, all Fock state can be  denoted with just three
numbers as $| N_{\bar{u}},N_{\bar{d}},N_{g}>$.

 In order to derive the state density $\rho_{|N_{\bar{u}},N_{\bar{d}},N_{g}>}$
 we should know the probability of states shifting.
  We introduce  the rate $f_{q\rightarrow qg}$ as a quark splitting ability
factor, there are $2N_{\bar{u}}+2N_{\bar{d}}+3$ quarks(including
antiquarks) in the initial state, so the NSSP of
$|N_{\bar{u}},N_{\bar{d}},N_{g}>\rightarrow|N_{\bar{u}},N_{\bar{d}},N_{g}+1>$
is
\begin{eqnarray}
(2N_{\bar{u}}+2N_{\bar{d}}+3)f_{q\rightarrow qg}.
\end{eqnarray}
 We also introduce the splitting rate $f_{g\rightarrow q\bar{q}}$ and $f_{g\rightarrow gg}$, so
 the NSSP of
$|N_{\bar{u}},N_{\bar{d}},N_{g}>\rightarrow|N_{\bar{u}},N_{\bar{d}}+1,N_{g}-1>$
and
$|N_{\bar{u}},N_{\bar{d}},N_{g}>\rightarrow|N_{\bar{u}}+1,N_{\bar{d}},N_{g}-1>$
is
\begin{eqnarray}
N_{g}f_{g\rightarrow q\bar{q}},
\end{eqnarray}
and the NSSP of
$|N_{\bar{u}},N_{\bar{d}},N_{g}>\rightarrow|N_{\bar{u}},N_{\bar{d}},N_{g}+1>$
is
\begin{eqnarray}
N_{g}f_{g\rightarrow gg}.
\end{eqnarray}

Now, we consider the time-reversal process and assume those fusion
rates
\begin{eqnarray}
f_{qg\rightarrow q}=f_{q\rightarrow qg},\nonumber
\\ f_{q\bar{q}\rightarrow g}=f_{g\rightarrow
q\bar{q}},\nonumber
\\ f_{gg\rightarrow
g}=f_{g\rightarrow gg}\nonumber
\end{eqnarray}
for time-reversal invariance.

Hence, the NSSP of
$|N_{\bar{u}},N_{\bar{d}},N_{g}>\rightarrow|N_{\bar{u}},N_{\bar{d}},N_{g}-1>$
is
\begin{eqnarray}
(2N_{\bar{u}}+2N_{\bar{d}}+3)N_gf_{qg\rightarrow
q}+\frac{N_g(N_g-1)}{2}f_{gg\rightarrow g},
\end{eqnarray}
the NSSP of
$|N_{\bar{u}},N_{\bar{d}},N_{g}>\rightarrow|N_{\bar{u}}-1,N_{\bar{d}},N_{g}+1>$
is
\begin{eqnarray}
(N_{\bar{u}}+2)N_{\bar{u}}f_{q\bar{q}\rightarrow g},
\end{eqnarray}
the NSSP of
$|N_{\bar{u}},N_{\bar{d}},N_{g}>\rightarrow|N_{\bar{u}},N_{\bar{d}}-1,N_{g}+1>$
is
\begin{eqnarray}
(N_{\bar{d}}+1)N_{\bar{d}}f_{q\bar{q}\rightarrow g}.
\end{eqnarray}
We can see that the probability of $u\bar{u}$ annihilation is larger
than $d\bar{d}$ annihilation in all of the proton states because of
valence quark asymmetry. This is the origin of the sea quark flavor
asymmetry.

It is assumed that all the splitting and fusion rates are the same
in the previous papers~\cite{Zhang1}. If we get all the
non-normalized state-shift probabilities $c_{ij}$, the state
densities can be derived out if the particle numbers
$N_{\bar{u},\bar{d},g}$ are finite. We set an artificial limit
$N_{\bar{u},\bar{d},g}\leq N_{max}$ and solve the finite linear
equations. The numeric state densities are then derived. The sea
quark flavor asymmetry can be written as:
\begin{eqnarray}
[\bar{d}-\bar{u}]=\sum_{\bar{u},\bar{d},g}(N_{\bar{d}}-N_{\bar{u}})\rho_{|N_{\bar{u}},N_{\bar{d}},N_{g}>}.
\end{eqnarray}
The sea quark flavor asymmetry converges to 0.133 when $N_{max}$
increases. The result is consistent with experiment
data~\cite{nmc,NA51,HERMES,E866a,E866}. Some subsequent
works~\cite{Singh,Alberg} followed the kinetic equilibrium principle
to study the spin of nucleons and the parton distributions in the
proton and pion, and obtained quite good results agreeing with the
corresponding experimental values.

However, in the previous works, we assumed that all the splitting
rates are the same as $f_{q\to qg} = f_{g\to q \bar q}= f_{g\to g
\bar g} \equiv 1$  and did not estimate the ``error bound'' caused
by the assumption. As we can imagine, if the splitting-rates vary in
different orders of magnitude, the convergence of flavor asymmetry
will be bad. It is necessary to solve large $N_{max}$ linear
equations. So we need a convenient numerical method to explore the
effects of different splitting-rates and to study more complex
hadronic states.

\section{Monte Carlo simulation approach of a kinetic equilibrium ensemble}
Monte Carlo simulation also can give the numeric state densities
instead of solving algebraic equations even when the number of
states is infinite. Here, we want to explain some details about the
Monte Carlo evolution on kinetic equilibrium and prove the
equivalence between Monte Carlo evolution approach and solving
algebraic equations. Let us start with an arbitrary initial state
$|i>$, and then let it make a possible shifting during each unit
step. The probability of the state $|i>$ shifting to $|j>$ is
$c_{ij}/C_0$. Here, $C_0$ is an arbitrary large constant we
introduced to ensure that the total shifting probability for each
prior state is less than $1$. It is required that $C_0>\sum_{j\neq
i} c_{ij}$ for all prior states $|i>$, so the probability of staying
in the prior state $|i>$ is
\begin{eqnarray}
1-\sum_{j\neq i} c_{ij}/C_0.
\end{eqnarray}
The state evolves step-by-step as random walk, and we record the
number of iteration steps as $T_i$ while the state $|i>$ is
emerging. And after a large number of iteration steps $T$, the
normalized $|i>$ emerging probability is $ T_{i}/T$. For each step
while the state is $|i>$, the next step has the probability
$c_{ij}/C_0$ to be $|j>$. So there are the times $T_i c_{ij}/C_0$
of state shifting $|i>\rightarrow |j>$. Of course, other states also
can shift to $|j>$, meanwhile $|j>$ has chance to stay at $|j>$.
That means the number of those steps  $|j>$ emerging  should be
\begin{eqnarray}
T_j=\sum_{i\neq j} c_{ij}/C_0 T_i+(1-\sum_{i\neq j} c_{ji}/C_0)T_j.
\end{eqnarray}
The equation can be reduced to
\begin{eqnarray}
\sum_{i\neq j} c_{ij} T_i=\sum_{i\neq j} c_{ji}T_j.
\end{eqnarray}
The equation is independent of the constant $C_0$. The value of
$C_0$ only determines the number of iteration steps needed to arrive
at the equilibrium state after starting from an arbitrary initial
state. We can find the above equation is just the kinetic
equilibrium equation (\ref{eq1}), if we consider that the normalized
$|i>$ emerging probability $ T_{i}/T$ is equivalent to the state
density as
\begin{eqnarray}
 T_i/T=\rho_i.
\end{eqnarray}

And we also have the sum condition
\begin{eqnarray}
\sum_{i} T_i=T,
\end{eqnarray}
which is equal to the normalization condition Eq(\ref{eq2}). Hence,
we proved the equivalence of the Monte Carlo simulation approach and
solving algebraic equations.

The Monte Carlo simulation approach provides a powerful method for
solving kinetic equilibrium  ensemble problems. This method is
error-controllable and very useful especially on complex multistate
systems, such as the applications to other hadrons in the following
sections. We gain the same value of the sea quark flavor asymmetry
$0.132\pm 0.02$ in the proton as expected. Here, the error bar $\pm
0.02$ is the standard deviation of results with different random
number series, and the deviation will decrease when computing time
increases.

\section{dynamics-nonsensitive sea quark flavor asymmetry in proton } The
fusion rate should be the same as the splitting rates for a
time-reversal process. In other words, the evolution in the proton
should be time-reversal invariant. But there is no principle
requires that the quark and gluon splitting evolution abilities of
$g\rightarrow q\bar{q}(gg)$ and $q\rightarrow qg$ are equal.
Therefore we should introduce three splitting-rates $f_{q\rightarrow
qg}$,$f_{g\rightarrow q\bar{q}}$ and $f_{g\rightarrow gg}$ to
represent the quark and gluon splitting evolution abilities which
are determined by the dynamics of quarks and gluons. Each rate
enhances the corresponding splitting or fusion evolution
probability. In previous works, we assumed that all the
splitting-rates are the same to be $f_{q\to qg} = f_{g\to q \bar q}=
f_{g\to g \bar g} \equiv 1$ and did not estimate the ``error band''
caused by the assumption. In the present work, we introduced a
numerical Monte Carlo approach. This new method is easy to apply to
complex systems, and it is easy to put the variable splitting rates
in evolutions and calculate the deviation caused by them.

In the above section, we can see that the state densities or results
are independent of the constant $C_0$. The numerical value of
$f_{g\to q\bar q}$, for example, is input as $f_{g\to q\bar q}/C_0$.
Therefore the result does not depend on the absolute value of
$f_{g\to q\bar q}$. It means that the sea quark asymmetry does not
depend on the absolute values of those splitting rates. Only two
ratios between three splitting rates will affect the state densities
and the value of sea quark flavor asymmetry. So, we can fix the rate
$f_{q\rightarrow qg}\equiv 1$, and vary the other two ratios
$f_{g\rightarrow q\bar{q}}/f_{q\rightarrow qg}$ and $f_{g\rightarrow
gg}/f_{q\rightarrow qg}$ as two parameters in the model.

\begin{table}[h]
\caption{The values of sea quark asymmetry for different ratios of
splitting rates}\label{table1}
\begin{tabular}{ccccccccc}
 \tableline
$[\bar{d}-\bar{u}]\times 100$& & & $f_{g\rightarrow q\bar{q}}/f_{q\rightarrow qg}$ & &\\
${f_{g\rightarrow gg}}/f_{q\rightarrow qg}$  & 100 & 10&1&0.1 &0.01 &0.001\\
 \tableline
   0&$\;\;\;123\pm 2 \;\;\;$&$\;\;\;124\pm 2 \;\;\;$&$ \;\;\;124\pm 2 \;\;\;$&$\;\;\;124\pm 3 \;\;\;$&$\;\;\;125\pm 3 \;\;\;$&$\;\;\;126\pm 6 \;\;\;$\\
   1&$\;\;\;131\pm 2 \;\;\;$&$\;\;\;132\pm 2 \;\;\;$&$ \;\;\;132\pm 2 \;\;\;$&$\;\;\;134\pm 3 \;\;\;$&$\;\;\;135\pm 3 \;\;\;$&$\;\;\;136\pm 6 \;\;\;$\\
  2&$137\pm 2$&$138\pm 3$&$140\pm 3$&$140\pm 4$&$141\pm 3$&$141\pm 6$\\
  5&$150\pm 2$&$152\pm 3$&$153\pm 3$&$154\pm 3$&$156\pm 4$&$156\pm 7$\\
 10&$161\pm 3$&$163\pm 3$&$164\pm 4$&$164\pm 3$&$165\pm 5$&$166\pm 8$\\
  \tableline
100&$179\pm 4$&$180\pm 4$&$180\pm 4$&$180\pm 3$&$181\pm 5$&$182\pm 9$\\
  \tableline
 \end{tabular}
\end{table}

In Table \ref{table1}, the values of sea quark asymmetry for
different ratios of splitting rates are listed. The previous result
$0.132\pm 0.02$ is reproduced when $f_{g\rightarrow
q\bar{q}}/f_{q\rightarrow qg}=f_{g\rightarrow gg}/f_{q\rightarrow
qg}=1$.

 From Table \ref{table1}, we can see that the asymmetry value $[\bar{d}-\bar{u}]$ is not
 sensitive to the model
parameter  $f_{g\rightarrow q\bar{q}}/f_{q\rightarrow qg}$, it is
almost fixed when $f_{g\rightarrow q\bar{q}}/f_{q\rightarrow qg}$
varies in  a very large range over five order of magnitudes. We also
can find that the values of asymmetry are always larger than $0.123$
whatever the splitting rates vary in an arbitrary large range. It
reflects the principle of statistics contributes the dominant part
of sea quark flavor asymmetry. The asymmetry only has a variation
 $[\bar d-\bar u]=(0.12-0.16)$ which is within $30\%$ when $f_{g\to
gg}/f_{q\to qg} $ varies in  the range $0\le f_{g\to gg}/f_{q\to
qg}\le 10$, and still a small variation $[\bar d-\bar
u]=(0.12-0.18)$ even when $f_{g\to gg}/f_{q\to qg} $ varies in a
larger magnitude range $0\le f_{g\to gg}/f_{q\to qg}\le 100$. So the
effect brought from details of the dynamics is small and within the
bound of the experiments' uncertainty.

By now, we do not consider the probability of $g\rightarrow ggg$
 splitting and $ggg\to g$ recombination yet, because the probability is
suppressed by coupling constant and ``three-body'' splitting
kinematics. $g\rightarrow ggg$ can be regarded as two successive
$g\rightarrow gg$, and its effect is same as the effect of
increasing $f_{g\rightarrow gg}$, as we can see from Table
\ref{table3}. However, the rate of three-body splitting $g\to ggg$
must be much smaller than two-body splitting $g\to gg$ or $q\to qg$,
because the three-body phase space in perturbative QCD is suppressed
by a factor of 2-3 order of magnitudes comparing with the two-body
splitting. Though the parton splitting in hadrons is a
strong-coupling non-perturbative process, we believe that we still
can safely assume $f_{g\to ggg}/f_{q\to qg}\ll0.1$ which only causes
a very small enhancement as shown in Table \ref{table3}.  The effect
of the splitting $g\to ggg$ is thus negligible.

\begin{table}[h]
\caption{The values of sea quark asymmetry $[\bar{d}-\bar{u}]\times
100$ for different value of ${f_{g\rightarrow ggg}/f_{q\to qg}}$,
for ${f_{g\rightarrow qg}}=1,f_{q\rightarrow gg}=1,f_{g\rightarrow
q\bar{q}}=1$ and $f_{g\rightarrow ggg}=f_{ggg\to g}$ }
 \label{table3}
\begin{tabular}{ccccccc}\tableline
& & &${f_{g\rightarrow ggg}/f_{q\to
qg}}$&\\
\tableline
   0&0.1&0.2&0.4&0.6&0.8&1.0  \\
  $\;\;\;132\pm 2\;\;\;$&$\;\;\;135\pm 2\;\;\;$&$\;\;\;137\pm 2\;\;\;$&$\;\;\;142\pm 3\;\;\;$&$\;\;\;145\pm 3\;\;\;$&$\;\;\;148\pm 3\;\;\;$&$\;\;\;150\pm
  4\;\;\;$\\
  \tableline

\end{tabular}
\end{table}

 Because the effect of the splitting $g\to ggg$ and recombination $ggg\to g$ is
negligible and the asymmetry value of $[\bar{d}-\bar{u}]$ is almost
independent of the parameter  $f_{g\rightarrow
q\bar{q}}/f_{q\rightarrow qg}$, there is only one parameter
$f_{g\rightarrow gg}/f_{q\rightarrow qg}$ can vary the asymmetry.
This parameter is QCD relevant and it is the only input from
dynamics. If the parameter could be fixed by analysis of QCD, the
deviation on sea quark asymmetry caused by the details of dynamics
can be determined and the sea quark flavor asymmetry in proton is
predictable.

These two splitting vertices are QCD vertices and have the same
coupling constant. The splitting kinematics of $g\to gg $ and $q\to
qg $ are also similar. So,  the splitting rates of  $g\to gg $ and
$q\to qg $ should be in the same order of magnitude. The assumption
can be supported by the integrations of  Altarelli-Parisi(A-P)
splitting functions. Though these equations are valid in the
perturbative region and the parton splitting  in hadrons is a
nonperturbative process, the ratio of the total splitting rates is
still inspirational. The ratio parameter $f_{g\rightarrow
gg}/f_{q\rightarrow qg}$ can be heuristically ``derived'' from
Altarelli-Parisi splitting functions\cite{AP}.

  The A-P splitting functions are
\begin{eqnarray}
&&P(q\to q(z)g)=C_F \frac{1+z^2}{1-z} ,\nonumber \\
&&P(g\to g(z)g)=C_A[\frac{1-z}{z}+\frac{z}{1-z}+z(1-z)] ,\nonumber \\
&&P(g\to q(z)\bar q)=T_R[z^2+(1-z)^2], \nonumber
\end{eqnarray}
where the color factors $C_F=4/3$, $C_A=3$ and $T_R=1/2$.

 The integrations of A-P splitting
  functions are assumed to be the total probabilities of quarks and gluons
  splitting. So the splitting-rates directly to be:
\begin{eqnarray}
&&f_{q\to qg}=\int_{0}^{1-z_{min}}P(q\to q(z)g) dz ,\nonumber \\
&&f_{g\to gg}=\int_{z_{min}}^{1-z_{min}}P(g\to g(z)g) dz ,\nonumber \\
&&f_{g\to q \bar q}=\int_{0}^{1} P(g\to q(z)\bar q) dz .\nonumber
\end{eqnarray}

The rates $f_{q\to qg}$ and $f_{g\to gg}$ are logarithmic divergent
when the integration limit $ z_{min} \to 0 $, but fortunately  the
ratio between the two rates is not divergent, and thus we have the
model parameter

\begin{eqnarray}
&&\frac{f_{g\to gg}}{f_{q\to
qg}}=\frac{\int_{z_{min}}^{1-z_{min}}P(g\to g(z)g)
dz}{\int_{0}^{1-z_{min}}P(q\to q(z)g)dz} \rightarrow
\frac{C_A}{C_F}=\frac{9}{4}, \nonumber
\end{eqnarray}
when $z_{min}\to 0$. The ratio parameter is not sensitive to the
integration limit $ z_{min}$. For example, when $ z_{min}=0.1$, the
ratio is $2.01$ which is close to $9/4$. Such small deviation change
on parameter $f_{g\rightarrow gg}/f_{q\rightarrow qg}$ dose not have
effect on sea quark asymmetry. Considered the integration limit is
relative to $Q^2$ scale, then the model parameter $f_{g\rightarrow
gg}/f_{q\rightarrow qg}$  and sea-quark asymmetry are not sensitive
to $Q^2$ scale. We estimated the ratio parameter by the perturbative
A-P splitting functions, it is just the ratio of color factors. We
assume the parameter value is still similar in the nonperturbative
region.

  The nonsensitive parameter $f_{g\rightarrow q\bar q}/f_{q\rightarrow
  qg}$ also can be derived by above method. But, it is relevant to the integration limit
  or $Q^2$ scale. The dependence can be extracted as $ \frac{-0.075T_R}{C_F\log z_{min}}$ when $z_{min}$ is small on the order of magnitude and becomes zero when $z_{min}\to 0$.
  For example, the value of parameter $f_{g\rightarrow q\bar q}/f_{q\rightarrow
  qg}=0.005$ when $z_{min}= 10^{-6}$, and the value is not sensitive to the magnitude of $z_{min}$ or $Q^2$ scale because of its $\log z_{min}$ dependence. We can see from Table.\ref{table1}, the sea-quark asymmetry
  is not sensitive to this parameter even it is so small.

 As discussed above, the ratio $f_{g\rightarrow gg}/f_{q\rightarrow qg}$
  is almost fixed to ratio of color factors as $9/4$ and the asymmetry is independent of other details except the parameter $f_{g\rightarrow gg}/f_{q\rightarrow qg}$.
   Therefore we arrived at the following conclusion: after considering  the detail
of QCD especially the color factors, we can predict the sea quark
flavor asymmetry in proton is $0.142\pm0.03$. It is enhanced a
little compared to the value given in the previous papers. More
precise measurement of $[\bar d-\bar u]$ is needed to examine the
statistical balance model.

The $x$-dependent $\left[\bar d(x) - \bar u(x)\right]$ can be
derived from  deep inelastic scattering and Drell-Yan processes, and
$\int_0^1 \left[ \bar d(x) - \bar u(x)\right] dx $ is given by
extrapolating $\left[\bar d(x) - \bar u(x)\right]$ to $x\rightarrow
0$ and $x\rightarrow 1$. The sea quark asymmetry values from three
collaborations are listed in Table. \ref{table4}, they are all
consistent with the sea quark asymmetry value predicted above. The
value of E866 seems a little bit smaller compared to the prediction
value, but the $x$ range of the E866 measurement is narrow and the
uncertainty brought by extrapolating to small $x$ is out of control.
So, more precise measurements are needed to test the prediction.

\begin{table}[h]
  \caption{$\int \left[\bar d(x) - \bar u(x)\right] dx $ as determined by three
  experiments.  The range of the measurement is shown along with the
  value of the integral over all $x$ ($Q^2=54~\rm{GeV^2/c^2}$).}
  \label{table4}

    \begin{tabular}{ccc} \tableline
Experiment   & $x$ range       & $\;\;\;\int_0^1 \left[ \bar d(x) -
\bar
u(x)\right] dx \;\;\;$\\
\tableline
E866         & $\;\;\;0.015<x<0.35\;\;\;$  & $\;\;\;0.118 \pm 0.012\;\;\;$ \\
NMC          & $0.004<x<0.80$  & $0.148 \pm 0.039$ \\
HERMES       & $0.020<x<0.30$  & $0.16  \pm 0.03$ \\
\tableline
   \end{tabular}
\end{table}

\section{ sea quark flavor asymmetry  in mesons }
Because the sea quark asymmetry value is not sensitive to details of
dynamics and only depends on the parameter $f_{g\rightarrow
gg}/f_{q\rightarrow qg}$ which is almost fixed as $9/4$, then it
should not only work for the proton, but also for the mesons and
other baryons. We suppose the statistical model also has validity on
predicting sea quark asymmetry in other hadrons. M. Alberg , E. M.
Henley \cite{Alberg} and C.-B. Yang \cite{cbYang} derived the parton
distributions of pions according the statistical model, but the sea
quark  asymmetry
 is zero because of the same valence quark number in pions. While the valence quark numbers of the $u$ and $d$ quarks are different for the
 kaons, for example, $K^+(u\bar{s})$ has one $u$ valence quark and
 no  $d$ valence quark. The statistical balance model predicts the sea quark  asymmetry value
 $\bar{d}-\bar{u}=0.284$ in $K^+$, when $f_{g\rightarrow gg}/f_{q\rightarrow
qg}=9/4$. In the same way, the sea quark asymmetry value
 $[\bar{d}-\bar{u}]=-0.275$ in $K^0(d\bar{s})$ and $[d-u]=-0.275$ in
 $\bar{K^0}(\bar{d}s)$, $d-u=0.275$ in
 $K^-(\bar{u}s)$. These sea quark asymmetry values are also not sensitive to
dynamics as shown in Table \ref{table5}.

 \begin{table}[h]
\caption{The values of sea quark asymmetry $\bar{d}-\bar{u}$ in
$K^+$ for different split factors }\label{table5}
\begin{tabular}{c|ccccccc}
 \tableline
$[\bar{d}-\bar{u}]$& & $f_{g\rightarrow q\bar{q}}/f_{q\rightarrow qg}$ &\\
 \tableline
${f_{g\rightarrow gg}}/f_{q\rightarrow qg}$ &100 & 10&1&0.1 &0.01 \\
 \tableline
   0&$\;\;\;0.263 \;\;\;$&$ \;\;\;0.264\;\;\;$&$\;\;\;0.264 \;\;\;$&$\;\;\;0.264 \;\;\;$&$\;\;\;0.265 \;\;\;$\\
  0.1&$\;\;\;0.264 \;\;\;$&$ \;\;\;0.265 \;\;\;$&$\;\;\;0.265\;\;\;$&$\;\;\;0.266 \;\;\;$&$\;\;\;0.266 \;\;\;$\\
  1&$0.272$&$0.274$&$0.275$&$0.277$&$0.278$\\
  5&$0.296$&$0.300$&$0.303$&$0.304$&$0.305$\\
 10&$0.311$&$0.312$&$0.312$&$0.312$&$0.313$\\
  \tableline
 \end{tabular}
\end{table}

 We can see from Table \ref{table5} that the asymmetry $[\bar{d}-\bar{u}]$ is independent of  $f_{g\rightarrow
q\bar{q}}/f_{q\rightarrow qg}$ and varies in a small range
0.263-0.31 as $f_{g\rightarrow gg}/f_{q\rightarrow qg}$ varies in a
large range 0-10.

\section{ sea quark flavor asymmetry  in baryons }
We also use our statistical model to predict sea quark asymmetry for
baryons. In a previous paper\cite{lShao}, L. Shao \emph{et al}.
derived the octet baryons' sea quark asymmetry values by the  method
of solving
  linear equations. They give $[\bar{d}-\bar{u}]=0.41$ in
 $\Sigma^+(uus)$ and $[\bar{d}-\bar{u}]=0.276$ in
 $\Xi^+(uss)$. In this paper, we get the same number by the Monte Carlo
 approach. We can find that the sea quark asymmetry value in
 $\Xi^+(uss)$ is almost the same as the meson  $K^+(u\bar{s})$ because
 their $u$ and $d$ valence quark numbers are the same. So, in the statistical
 model, the $s$ valence quark number in the hadron has a negligible effect on
the $[\bar{d}-\bar{u}]$ sea quark asymmetry. We also find the sea
quark  asymmetry values in the octet baryons are not sensitive to
details of dynamics, they just depend on the valence quark numbers
in those baryons. The asymmetries $[\bar{d}-\bar{u}]$ in
$\Sigma^+(uus)$ and $\Xi^+(uss)$ are enhanced a little to be $0.42$
and $0.285$ when $f_{g\rightarrow gg}/f_{q\rightarrow qg}=9/4$.

Besides octet baryons, we also derived $\Delta$ baryons' sea quark
asymmetry value as:
\begin{eqnarray}
&&\bar{d}-\bar{u}=0.50 \;\; \;\;\rm{for} \;\;\Delta^{++}(uuu)\; ,\nonumber\\
&&\bar{d}-\bar{u}=0.14 \;\; \;\;\rm{for} \;\;\Delta^{+}(uud)\; ,\nonumber\\
&&\bar{d}-\bar{u}=-0.14 \;\;\;\;\rm{for} \;\; \Delta^{0}(udd)\; ,\nonumber\\
&&\bar{d}-\bar{u}=-0.50 \;\; \;\;\rm{for} \;\;\Delta^{-}(ddd)\;
,\nonumber
\end{eqnarray}

where, $f_{g\rightarrow gg}/f_{q\rightarrow qg}=9/4$. The sea quark
asymmetry in $\Delta^{+}(uud)$ is the same as in proton because of
their same $u$ and $d$ valence quark numbers. Of course, the
asymmetry in $\Delta^{0}(udd)$ is the same as in neutron.

We also derived exotic baryons' (pentaquark states) sea quark
asymmetry values as :
\begin{eqnarray}
&&\bar{d}-u=-0.14 \;\; \;\; \rm{for} \;\;\Phi^{--}(ssdd\bar{u}) ,\nonumber\\
&& d-\bar{u}=0.14 \;\; \;\; \rm{for}
\;\;\Phi^{-}(ssuu\bar{d}),\nonumber
\end{eqnarray}
where, the sea quark asymmetry values are the same as in the proton
because of their same $u(\bar{u}) $ and $d(\bar{d})$ valence quark
numbers.

If there is such a pentaquark state $X^{++}(uuud\bar{s})$, then its
sea quark asymmetry value would be $[\bar{d}-\bar{u}]=0.21$ derived
by the statistical model.

\section{Conclusions}
In the previous works in the statistical balance model, the sea
quark flavor asymmetry $[\bar d-\bar u]\equiv\int dx(\bar d(x)-\bar
u(x))$ in the proton was computed using the ``linear equations
method''. Because of the difficulty and limit of the linear
equations method, it is hard to apply the method to more complex
systems. It is also assumed that all the splitting-rates are the
same,  $f_{q\to qg} = f_{g\to q \bar q}= f_{g\to g \bar g} \equiv 1$
in the previous works, and the ``error band'' caused by the
assumption was not estimated. In the present work, we introduced a
numerical Monte Carlo approach. This new method is easy to apply to
complex systems, such as other mesons and baryons. We also
introduced the variable splitting rates representing details of the
dynamics and studied the dependence on them. We find the sea quark
flavor asymmetry in the proton is always larger than $0.123$
whatever the splitting rates vary over an arbitrary large range. It
reflects the statistics principle contributes the dominant part of
the asymmetry. The asymmetry is almost independent of the model
parameter $f_{g\to q \bar q}/f_{q\to qg} $ and only changes within
30\% when $f_{g\to gg}/f_{q\to qg} $ varies in the range $0-10$. So
the effect caused by details of the dynamics is small and within the
bound of the experiments' uncertainty. However, these two splitting
vertices are QCD vertices and have the same coupling constant. The
splitting kinematics of $g\to gg $ and $q\to qg $ are also similar.
So the splitting rates of  $g\to gg $ and $q\to qg $ should be in
the same order of magnitude. The assumption can be supported by the
integrations of  Altarelli-Parisi splitting functions. Though these
equations are valid in the perturbative region, one may
heuristically assume that the ratio of the total splitting rates
obtained from them holds approximately also in the nonperturbative
regime. The parameter $f_{g\to gg}/f_{q\to qg} $ can be fixed to the
ratio of color factors as $9/4$ by integrations of Altarelli-Parisi
splitting functions. According to the above reasons, we can conclude
that the prediction only from a statistics principle has an accuracy
$<30\%$. Or, in other words, the details of the dynamics only bring
less than 30\% effect. After considering the details of QCD
especially the color factors, the sea quark flavor asymmetry in
proton is enhanced to $0.142\pm0.03$ which is consistent with
present experimental measurements and can be tested by more precise
measurements.

  The sea quark asymmetries are not sensitively dependent on the details of
dynamics. The sea-quark flavor asymmetry derived only from statistic
principle contributes the dominant part of the asymmetry. It
strongly implies that the origin of the sea-quark flavor asymmetry
of hadrons is the asymmetry of valence quarks. We also applied this
Monte Carlo approach of statistical model to predict the sea quark
asymmetries in kaons, octet baryons $\Sigma$, $\Xi$, and $\Delta$
baryons, even in exotic pentaquark states. All these asymmetries
just only depend on the valence quarks number in those hadrons. The
sea-quark asymmetries for different $u$ and $d$ valence quark
numbers are listed in Table \ref{table6}. These values can confirm
the mechanism we proposed to explain the sea quark asymmetry in
proton. It can be observed from Table \ref{table6} that the sea
quark asymmetries are enhanced by the difference of corresponding
valence quark numbers and suppressed by the sum of valence quark
numbers. When the valence quark numbers $[u_v]>[d_v]$, the
sea-quarks $\bar u$ are easier to annihilated because of the
existence of more $u$ valence quarks and it leads the sea quark
asymmetry. On the other hand, the larger total number of valence
quark $[u_v+d_v]$ suppresses the relative difference of valence
quarks and weakens the sea quark asymmetries even if $[u_v-d_v]$
remains the same. These sea quark asymmetries for hadrons, except
the proton, are listed purely for theoretical interest, as it is not
known presently how to access this information in experiment.

\begin{table}[h]
  \caption{The sea-quark asymmetry values for
different $u,d$ valence quark numbers, $f_{g\rightarrow
gg}/f_{q\rightarrow qg}=9/4$.}
  \label{table6}

\begin{tabular}{c|ccccc}
 \tableline
asymmetry values& & $u$ valence quark number \\
 \tableline
$d$ valence quark number  &0 & 1&2&3  \\
 \tableline
   0&$\;\;\;0 \;\;\;$&$ \;\;\;0.284(K^+,\Xi^0) \;\;\;$&$\;\;\;0.42(\Sigma^+) \;\;\;$&$\;\;\;0.50(\Delta^{++}) \;\;\;$\\
  1&$\;\;\;-0.284(K^{0},\Xi^{-}) \;\;\;$&$ \;\;\;0(\Lambda^0,\Sigma^0) \;\;\;$&$\;\;\;0.14(P,\Delta^+,\Phi^-)\;\;\;$&$\;\;\;0.21(uuud\bar{s}) \;\;\;$\\
  2&$-0.42(\Sigma^-)$&$-0.14(N,\Delta^0,\Phi^{--})$&$0(\Theta^+,\Theta_c)$&$0.07(uuudd\bar{s}\bar{s})$\\
  3&$-0.50(\Delta^-)$&$-0.21(dddu\bar{s})$&$-0.07(uuddd\bar{s}\bar{s})$&$0$\\
  \tableline
 \end{tabular}
\end{table}

{\bf Acknowledgment:} This work of B.~Z. is supported by the
National Science Foundation of China under Grant No. 10705017 and
11075086. Y.J. Z. is supported by Liaoning Education Office
Scientific Research Project (2008288)


\begin{thebibliography}{10}
\bibitem{Gottfried}
K. Gottfried, Phys. Rev. Lett. \textbf{18}, 1174 (1967).




\bibitem{nmc}
New Muon Collaboration, P. Amaudruz {\it et al}., Phys. Rev. Lett.
\textbf{66}, 2712 (1991); M. Arneodo {\it et al}., Phys. Rev. D
\textbf{50}, R1 (1994).
\bibitem{NA51}  NA51 Collaboration, A. Baldit {\it et al.}, Phys. Lett. {\bf B 332}, 244 (1994).

\bibitem{HERMES} HERMES Collaboration, K. Ackerstaff {\it et al.}, Phys. Rev. Lett. {\bf 81}, 5519 (1998).

\bibitem{E866a} FNAL E866/NuSea Collaboration, E.A. Hawket {\it et al.},
Phys. Rev. Lett. {\bf 80}, 3715 (1998)

\bibitem{E866} FNAL E866/NuSea Collaboration, R.S. Towell {\it et al.},
Phys. Rev. {\bf D 64}, 052002 (2001).



\bibitem{FlavorAsymmetry}
 S. Kumano, Phys. Rep. {\bf 303}, 183
(1998);

\bibitem{Speth} J.P.Speth and A.W.Thomas,
Adv. Nucl. Phys.  {\bf 24}, 93 (1997).

\bibitem{Kumano}S.Kumano, Phys. Rep. 303, 183 (1998).

\bibitem{Melnitchouk} W. Melnitchouk, J. Speth, A.W. Thomas, Phys. Rev. D {\bf  59},
014033(1998).

\bibitem{Nikolaev}N.N.Nikolaev et al., Phys. Rev. D
{\bf 60 }, 014004 (1999).

\bibitem{Henley}M.Alberg, E.M.Henley and G.A.Miller, Phys.
Lett. {\bf B 471}, 396 (2000).

\bibitem{Magnin} J. Magnin, H.R. Christiansen, Phys. Rev. D {\bf  61},054006 (2000).

\bibitem{Garvey}
G.T. Garvey and J.-C. Peng, Prog. Part. Nucl. Phys. {\bf 47}, 203
(2001).

\bibitem{Pasquini} B. Pasquini, S. Boffi, Nucl. Phys. A {\bf 782  }, 86(2007).


\bibitem{Eichten} E.J. Eichten, I. Hinchliffe, C. Quigg, Phys. Rev. D {\bf 45 },2269 (1992).
\bibitem{Cheng} T.P. Cheng, L.-F. Li, Phys. Rev. Lett.  {\bf 74 },2872 (1995).
\bibitem{Brown}G.E. Brown, M. Rho, Phys. Rep. {\bf 363 },85 (2002).
\bibitem{Ding} Y. Ding, R.-G.
Xu, B.-Q. Ma, Phys. Lett. {\bf B 607}, 101 (2005); Y. Ding, B.-Q.
Ma, Phys. Rev. D {\bf 73}, 054018  (2006).

\bibitem{Pobylitsa}P.~V.~Pobylitsa, M.~V.~Polyakov, K.~Goeke, T.~Watabe and C.~Weiss,
Phys.\ Rev.\  D {\bf 59}, 034024 (1999) [arXiv:hep-ph/9804436].


\bibitem{Zhang1}  Y.-J. Zhang, B.Zhang, and B.-Q. Ma, Phys. Lett. {\bf B 523}, 260 (2001);
Y.-J. Zhang, W.-Z. Deng, B.-Q. Ma, Phys.Rev. D \textbf{65} 114005
(2002).


\bibitem{Fiel77} R. D. Field and R. P. Feynman, Phys. Rev. D {\bf 15}, 2590
(1977).

\bibitem{Ross79} D. A. Ross and C. T. Sachrajda, Nucl. Phys. B {\bf 149}, 497
(1979).

\bibitem{Stef97} F. M. Steffens and A. W. Thomas, Phys. Rev. C {\bf 55}, 900
(1997).



\bibitem{Singh}J.P. Singh and Alka Upadhyay,
J.Phys. \textbf{G30}, 881 (2004).

\bibitem{AP}G. Altarelli and G. Parisi, Nucl. Phys. B {\bf 126} (1977), 298.


\bibitem{Alberg} Y.-J. Zhang, B.-S. Zou, and L.-M. Yang, Phys. Lett. {\bf B 528}, 228 (2002);
M. Alberg and E.M. Henley, Phys. Lett. {\bf B 611}, 111 (2003).

\bibitem{cbYang} C.-B. Yang, Chin.Phys.Lett.20:821-824,(2003).

\bibitem{lShao} Lijing Shao, Yong-Jun Zhang, Bo-Qiang Ma,  Phys. Lett. {\bf B 686 }, 136(2010).

\end{thebibliography}
\end{document}